An arXiv paper flawed due to journal artist's error

# Note on: "Considering the Case for Biodiversity Cycles: Reexamining the Evidence for Periodicity in the Fossil Record" by Lieberman and Melott, arXiv preprint 0704.2896

Mensur Omerbashich

*Physics Department, Faculty of Science, University of Sarajevo, Zmaja od Bosne 33, Sarajevo, Bosnia*
*Phone +387-63-817-925, Fax +387-33-649-359, E-mail: momerbasic@pmf.unsa.ba; CC: omerbashich@yahoo.com*

Lieberman and Melott built their recent arXiv preprint 0704.2896 on my published paper and (a preprint of) a subsequent comment by Lieberman's associate Cornette. But had this group waited for the Cornette's comment to actually appear in print together with the expected Reply, they would have learned that his comment exposes Cornette's confusion that likely was due to journal misprint of my figure. Thus 0704.2896 is baseless. Despite receiving the extended Reply with Errata, these authors still fail to recognize that detrending of paleontological records—which they erroneously promote as a must—is an arbitrary rather than a universal operation.

[1] recently demonstrated that the Sepkoski compendium (world's most complete fossil record) offers no evidence for new life cycles, contrary to the claims made by [2]. The results of [1] were questioned by [3] that conversely was dismissed by [4] citing a discolored figure as a likely "basis" for [3]. This misprint could have had resulted in the ignorance displayed in [3] that has thus demonstrably confused my power spectra for my variance spectra, misunderstood the accompanying statistics, as well as exhibited a critical lack of expertise in Gauss-Vaníček spectral analysis; see [4] for details. This renders the Comment [3] entirely irrelevant for [1], and by the same token for any other considerations as well. Both [3] and [4] are to appear in print next month.

In the meantime however, a rather complex paper [5] has been built on [1] and [3] ("*Our work builds on…*" p.3 line 21 [4]). Presumably, a preprint of [3] was sent to the authors of [5] in advance of publication of [3]. Though it may appear a standard conduct to reference previous work even prior to publication, still neither was [3] a usual type of a paper (it was a Comment, so it was only normal to look for a Reply too) nor was [3] non-crucial for [5].

The hurry with which the authors of [5] based a complex and a hard-to-understand paper like [5] chiefly on *scientific hear-say* (Comments rarely appear in print without a Reply!) is surprising. Furthermore, nowhere (say, in form of their preprint's updated versions) do these authors cite [4] although its extended version with Errata has been sent to them as a courtesy after they posted [5] on arXiv. Thus the authors of [5] exhibited poor scholarship by not waiting for an expected Comment to appear in print together with the respective Reply, but also by ignoring the Reply and associated Errata a posteriori too.

## Conclusion

By ignoring what's obvious, the authors of [5] insist that the detrending of paleontological records— which they erroneously promote in [5] as a must—is a universal instead of an arbitrary operation. They do this in spite of to data inherent incompleteness, a detrender function selection, or any adverse affects that an unspecified combination of the two can have on the data and spectra alike. Therefore, I believe that [5] ought to be withdrawn.